\documentclass[10pt]{article}

\usepackage{amsmath}
\usepackage{amssymb}
\usepackage{graphicx}
\usepackage{psfrag}
\usepackage{color}
\usepackage{sistyle}

\newcommand{\drawat}[3]{\makebox[0pt][l]{\raisebox{#2}{\hspace*{#1}#3}}}

\definecolor{linkcolor}{rgb}{0,0,1}

\usepackage[colorlinks=true, pdfstartview=FitV, linkcolor= linkcolor, citecolor= linkcolor, urlcolor= linkcolor, hyperindex=false]{hyperref}

\begin{document}

\title{\bf Direct measurement of spatial modes of a micro-cantilever from thermal noise}

\author{
Pierdomenico Paolino\footnote{present address: Universit\'e de Lyon, LTDS, \'Ecole Centrale Lyon, CNRS UMR 5513, Ecully, France} \\
\emph {\small Universit\'e de Lyon, Laboratoire de physique, \'Ecole Normale Sup\'erieure de Lyon} \\
\emph {\small CNRS UMR 5672, Lyon, France} \\
Bruno Tiribilli \\
\emph {\small Istituto dei Sistemi Complessi, Consiglio Nazionale delle Ricerche, Sesto Fiorentino} \\
\emph {\small Firenze, Italy} \\
Ludovic Bellon\footnote{corresponding author : ludovic.bellon@ens-lyon.fr} \\
\emph {\small Universit\'e de Lyon, Laboratoire de physique, \'Ecole Normale Sup\'erieure de Lyon} \\
\emph {\small CNRS UMR 5672, Lyon, France} \\
}

\date{}

\maketitle

\begin{abstract}
Measurements of the deflection induced by thermal noise have been performed on a rectangular atomic force microscope cantilever in air. The detection method, based on polarization interferometry, can achieve a resolution of $\SI{10^{-14}}{ m/\sqrt{Hz}}$ in the frequency range $\SI{1}{kHz}-\SI{800}{kHz}$. The focused beam from the interferometer probes the cantilever at different positions along its length and the spatial modes' shapes are determined up to the fourth resonance, without external excitation. Results are in good agreement with theoretically expected behavior. From this analysis accurate determination of the elastic constant of the cantilever is also achieved.   
\end{abstract}

\vspace{5mm}

\noindent {\bf Notes}

\noindent Copyright 2009 American Institute of Physics. This article may be downloaded for personal use only. Any other use requires prior permission of the author and the American Institute of Physics.

\noindent This article appeared in \emph{Journal of Applied Physics} {\bf 106}, 094313 (2009) and may be found at

\noindent \href{http://link.aip.org/link/?JAPIAU/106/094313}{http://link.aip.org/link/?JAPIAU/106/094313}

\newpage

\section{Introduction}

Atomic force microscopy (AFM) is currently used in a great variety of studies involving small forces measurement \cite{2005butt} including unfolding of protein \cite{1999Fisher,1999Carrion-Vazquez},
 probing the structure of biological membranes \cite{2009Frederix} and monitoring the mechanical response of living cells \cite{2007Radmacher, 2008SBRANA} as well as Micro-Electro-Mechanical Systems (MEMS) and other nanotechnological devices \cite{2008Bhushan-b,Lavrik-2004}.

All those applications exploit the great accuracy in measuring the cantilever deflection offered by AFM and converting this measurement in units of force assuming the cantilever behaves like a spring with known stiffness. Manufacturers often specify the spring constant of their cantilevers in a wide range of values, mainly because of the great uncertainties in the dimensions, particularly the thickness, resulting from the fabrication process. To overcome this problem several techniques have been proposed to calibrate cantilever spring constant \cite{2005butt,cook,levy,1993hutter,sader-1999}. The reader is referred to the work of Burnham and co-workers \cite{Burnham-2003} and the references therein for a comparative summary of the different techniques.

One of the first and still most commonly used is the so called thermal calibration method \cite{1993hutter} based on the measurement of the vibration amplitude of the free end of a cantilever exited by thermal noise. The first peak of the thermal noise spectrum is related back to the spring constant of the cantilever modeled as an harmonic oscillator. In a more accurate model, Butt and Jaschke \cite{butt} introduced a correction factor deduced from the Euler-Bernoulli description of the flexural dynamic of a free-clamped beam.

In this work we measure thermal noise spectrums of the cantilever deflection along its length, and compare the rms amplitudes of the first four modes of vibration as a function of spatial position with the eigenmodes of the Euler-Bernoulli model. Furthermore, we present an extention of the thermal noise calibration method for the spring constant based on the multi-mode measurement.

\section{Theoretical background}

\begin{figure}[ht]
\begin{center}
\includegraphics{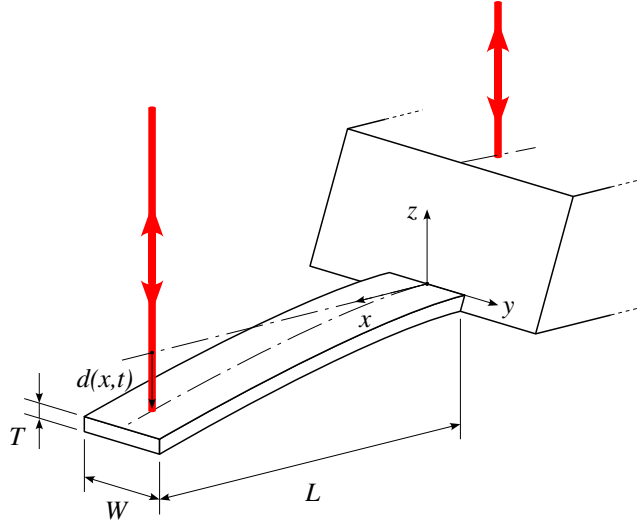}
\end{center}
\caption{The thermal fluctuations of deflection $d(x,t)$ of a rectangular cantilever (BS-Cont-GB-G) are measured with a differential interferometer : the optical path difference between the reference beam, reflecting on the chip holding the cantilever, and the sensing beam, focused on the cantilever, directly gives a spatially resolved and calibrated measurement of deflection $d$ \cite{patent-Bellon}. The whole cantilever can be probed by displacing the beams along its length ($x$ axis). Length $L$, thickness $T$ and width $W$ of the cantilever are indicated on the figure.}
\label{Fig:cantilever}
\end{figure}

Let us first recall the main lines of Butt and Jaschke's work \cite{butt} to interpret our measurements. The cantilever is sketched on Fig.~\ref{Fig:cantilever}. Its length $L$ is supposed to be much larger than its width $W$, which itself is much larger than its thickness $T$. We will limit ourself in this study to the flexural modes of the cantilever: the deformations are supposed to be only perpendicular to its length (along axis $z$ of Fig.~\ref{Fig:cantilever}) and uniform across its width. These deformations can thus be described by the deflection $d(x,t)$, $x$ being the spacial coordinate along the beam, and $t$ the time. The generic solution of the Euler-Bernoulli equation can be expressed as follows:
\begin{equation}\label{EBsolution}
          d(x,t)=\sum^{\infty}_{n=1}d_{n}(t)\phi_{n} \left(\frac{x}{L}\right)
\end{equation}
in which the spatial solutions are in the form
\begin{equation}\label{Spatial}
	\phi_{n} \left(\frac{x}{L}\right)= \left[\cos\left(\alpha_{n}\frac{x}{L}\right)-\cosh\left(\alpha_{n}\frac{x}{L}\right)\right]-\frac{\cos(\alpha_{n})+\cosh(\alpha_{n})}{\sin(\alpha_{n})+\sinh(\alpha_{n})}\left[\sin\left(\alpha_{n}\frac{x}{L}\right)-\sinh\left(\alpha_{n}\frac{x}{L}\right)\right]
\end{equation}
where the $\alpha_{n}$ satisfy the relation
\begin{equation}
          1+\cos(\alpha_{n}) \cosh(\alpha_{n})=0 
\end{equation}
which leads to  $\alpha_{1}=1.875$, $\alpha_{2}=4.694$, \ldots , $\alpha_{n} \approx(n-1/2)\pi$.
The amplitude $d_{n}(t)$ of each modes are governed by harmonic oscillator equations, with spring constants $k_{n}$, mass $m$ and resonance frequencies $f_{n}$:
\begin{equation}\label{kn}
          k_{n}=\frac{\alpha_{n}^{4}}{3}k_{c}=\frac{\alpha_{n}^{4}}{3}\frac{EWT^3}{4L^3} 
\end{equation}
\begin{equation}
          m=m_{c}=\rho L W T
\end{equation}
\begin{equation}\label{fn}
          f_{n}=\frac{1}{2 \pi}\sqrt{\frac{k_{n}}{m}}
\end{equation}
where $k_{c}$ and $m_{c}$ are the static stiffness and mass of the cantilever, and $E$ and $\rho$ are the Young's modulus and density of its material.

We furthermore consider, under the hypothesis of thermal equilibrium, that the thermal noise driven deflection follows the equipartition theorem and each resonance mode can be considered as an independent harmonic oscillator \cite{butt} with mean quadratic fluctuations $\left\langle d_{n}^2\right\rangle$:
\begin{equation}\label{eq:equirepartition} 
          \frac{1}{2}k_{B}T =\frac{1}{2}k_{n}\left\langle d_{n}^2\right\rangle
\end{equation}          
where $\left\langle\ \right\rangle$ represents time average.

\section{Experimental methodology and results}

In our experiment, we measure both the amplitude of the thermal noise distibution among the modes $\left\langle d_{n}^2\right\rangle$ and their spatial shape $\phi_{n} \left(x/L\right)$. We use gold coated BudgetSensors Atomic Force Microscopy (AFM) cantilevers (Cont-GB-G). They present a nominal rectangular geometry: $L=\SI{450}{\mu m}$ long, $W=\SI{50}{\mu m}$ wide and $T=\SI{2}{\mu m}$ thick, with a $\SI{70}{nm}$ gold layer on both sides. We observe the fluctuation of the cantilever deflection induced by thermal noise. The measurement is performed with a home made interferometric deflection sensor \cite{patent-Bellon, Paolino:afm}, inspired by the original design of Schonenberger \cite{1989Schonenberger} with a quadrature phase detection technique \cite{2002Bellon}: the interference between the reference laser beam reflecting on the chip of the cantilever \cite{patent-Bellon} and the sensing beam on the cantilever gives a direct measurement of the deflection $d(x,t)$, with very high accuracy (see Fig.~\ref{Fig:cantilever}).

\begin{figure}[ht]
\begin{center}
\includegraphics{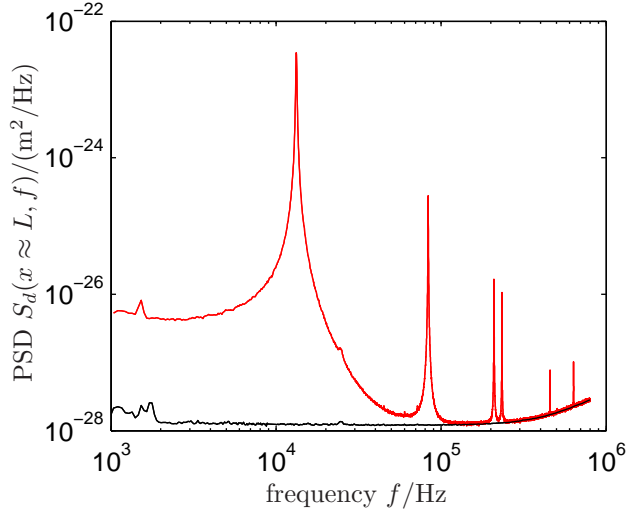}%
\drawat{-49mm}{1mm}{frequency $f/\SI{}{Hz}$}%
\drawat{-83mm}{17mm}{\rotatebox{90}{PSD $S_d(x\approx L,f)/(\SI{}{m^2/Hz})$}}%
\end{center}
\caption{Power Spectrum Density (PSD) $S_{d}(x\approx L,f)$ of thermal noise induced deflection (red curve) measured close to the free end of the cantilever as a function of frequency $f$ (log scale on both axis). The first 4 flexural modes and first 2 torsional modes are clearly above the background noise (bottom black line), measured with a rigid mirror.  With our interferometric setup, resolution better than $\SI{1.7e-14}{m/\sqrt{Hz}}$ can be achieved on the whole frequency range explored here ($\SI{1}{kHz}-\SI{800}{kHz}$).}
\label{Fig:spectrums}
\end{figure}

A first advantage of this technique is that it offers a calibrated measurement of deflection, without conversion factor from angle to displacement, as in the standard optical lever technique common in AFM. A second advantage of our detection system is a very low intrinsic noise, as illustrated in Fig.~\ref{Fig:spectrums} with the power spectrum density of a rigid mirror (bottom black line): the light intensities on the photo diodes are tuned exactly as during the measurement on the cantilever, but since the mirror is still, the measured spectrum reflects only the detection noise. This background noise is as low as $\SI{3e-28}{m^{2}/Hz}$ in the frequency range from $\SI{1}{kHz}$ to $\SI{800}{kHz}$, just $\SI{10}{\%}$ higher than the shot noise limit of our detection system. A third advantage is that the precision of the measurement is independent of the focused beam size on the cantilever, which was tuned as small as $\SI{10}{\mu m}$ to ensure good spatial resolution along the cantilever length.

When translating the focusing lens with respect to the cantilever, both reference and sensing laser beams are shifted along the chip and cantilever respectively \cite{patent-Bellon}. The chip's first structural mode is at very high frequency compared to that of the cantilever, hence it is considered as a rigid fixed mirror independently of the actual reference beam position on top of it. The measured interference signal is therefore only due to the thermal noise driven deflection of the cantilever. Fig.~\ref{Fig:spectrums} illustrates such a spectrum when the sensing beam is close to the free end. The signal to noise ratio is good enough to identify the first four flexural resonances (as well as the first two torsional resonances, which are visible due to imperfect centering of the spot laterally). The relative frequencies of the peaks should obey the following relation that directly derives from equations \ref{kn} and \ref{fn}:
\begin{equation}\label{Dispersion}
\frac{f_{n}}{f_{1}}=\frac{\alpha_{n}^2}{\alpha_{1}^2}
\end{equation}
The experimental ratios of resonance frequencies are in good agreement with the expected values (see table~\ref{table}).

\begin{table}[ht]
\begin{center}
\begin{tabular}{|c|c|c|c|c|c|}
\hline
Mode & $f_{n} /\SI{}{kHz}$ & $f_{n}/f_{1}$ & $\alpha_{n}^{2}/\alpha_{1}^{2}$ & $Q_{n}$ & $Q_{n}^{\textrm{Sader}}$ \\
\hline
1 & 14.046 &1 & 1 & 88 & 85 \\
\hline
2 & 87.921 & 6.26 & 6.27 & 231 & 243 \\
\hline
3 & 245.500 & 17.48 & 17.55 & 420 & 423 \\
\hline
4 & 479.970 & 34.17 & 34.40 & 680 & 601 \\
\hline
\end{tabular}
\end{center}
\caption{Measured and expected parameters for first 4 flexural modes.}
\label{table}
\end{table}

\begin{figure}[ht]
\begin{center}
\includegraphics{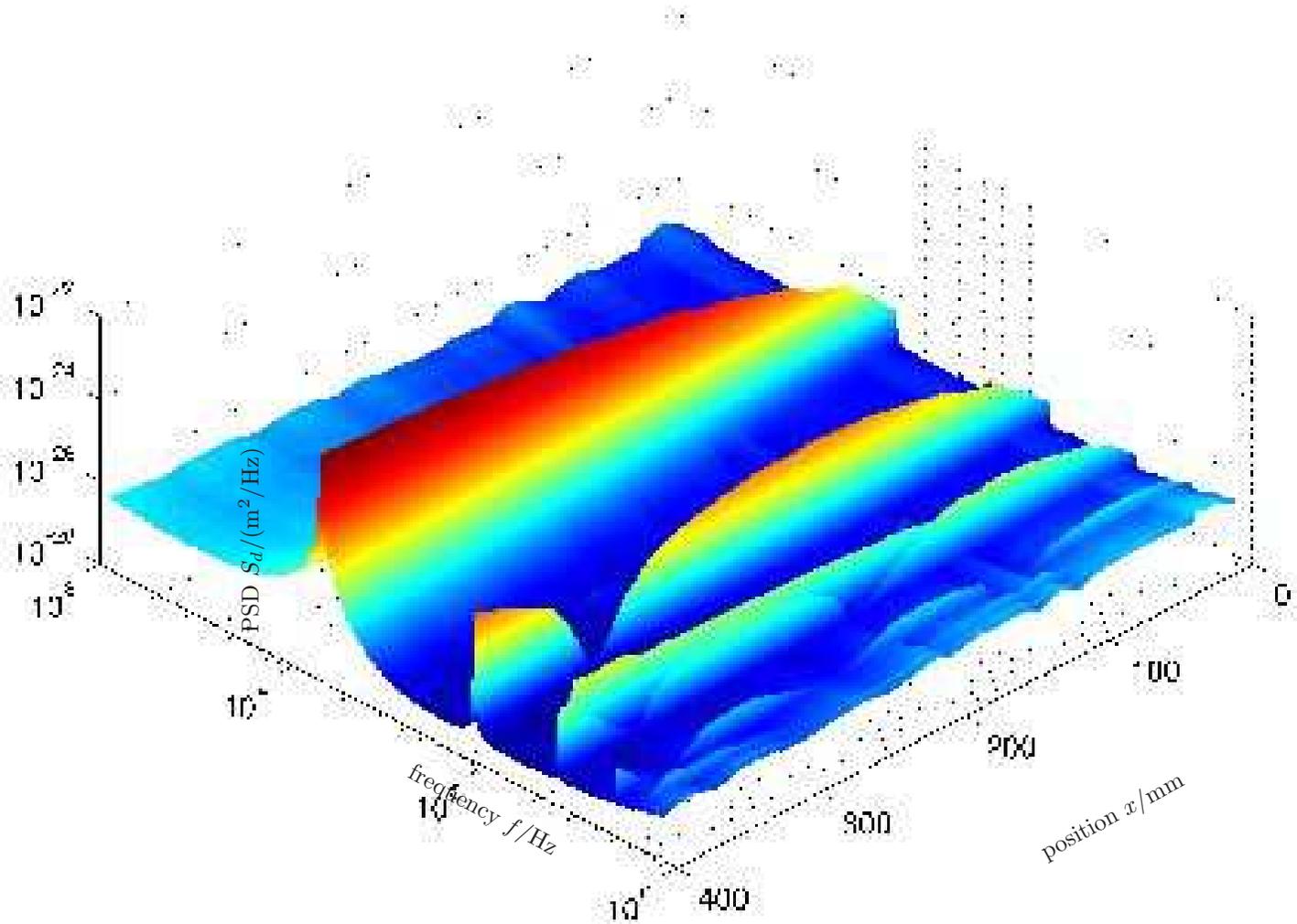}%
\drawat{-38mm}{10mm}{\rotatebox{28.3}{position $x/\SI{}{mm}$}}%
\drawat{-130mm}{22mm}{\rotatebox{-28.3}{frequency $f/\SI{}{Hz}$}}%
\drawat{-154mm}{43mm}{\rotatebox{90}{PSD $S_d/(\SI{}{m^2/Hz})$}}%
\end{center}
\caption{Power Spectrum Density (PSD) $S_{d}(x,f)$ of thermal noise induced deflection as a function of frequency $f$ and position $x$ along the cantilever. The first 4 normal modes are clearly visible, with a vanishing amplitude toward the clamped extremity of the mechanical beam and the nodes of each mode. Another vibration peak with no nodes is also visible close to the third mode, it is attributed to the first mode in torsion, but is not studied in this paper. }
\label{Fig:3D}
\end{figure}

A precision screw allows adjusting the horizontal position of the focusing lens and locating the sensing beam at different positions along the cantilever length while the reference beam is always on the chip \cite{patent-Bellon}. For each measurement, the $x$ position of the laser spot on the cantilever can be estimated from the image acquired by a simple CCD camera. The spot center in the images can be detected with $\pm\SI{1.4}{\mu m}$ accuracy on the cantilever. Measurements were repeated with approximative $\SI{15}{\mu m}$ steps. At every position we measure the deflection $d(x,t)$ produced by thermal excitation of the cantilever and evaluate the Power Spectrum Density (PSD) $S_{d}(x,f)$. The complete set of results is reported in Fig.~\ref{Fig:3D} as a 3D representation. The first four oscillation modes can be clearly seen with their respective number of nodes. Two further peaks can be noted, the first at about $\SI{220}{kHz}$ and a second, of smaller amplitude, at $\SI{790}{kHz}$, that we attribute to the first and second torsional modes. 

\begin{figure}[ht]
\begin{center}
\includegraphics{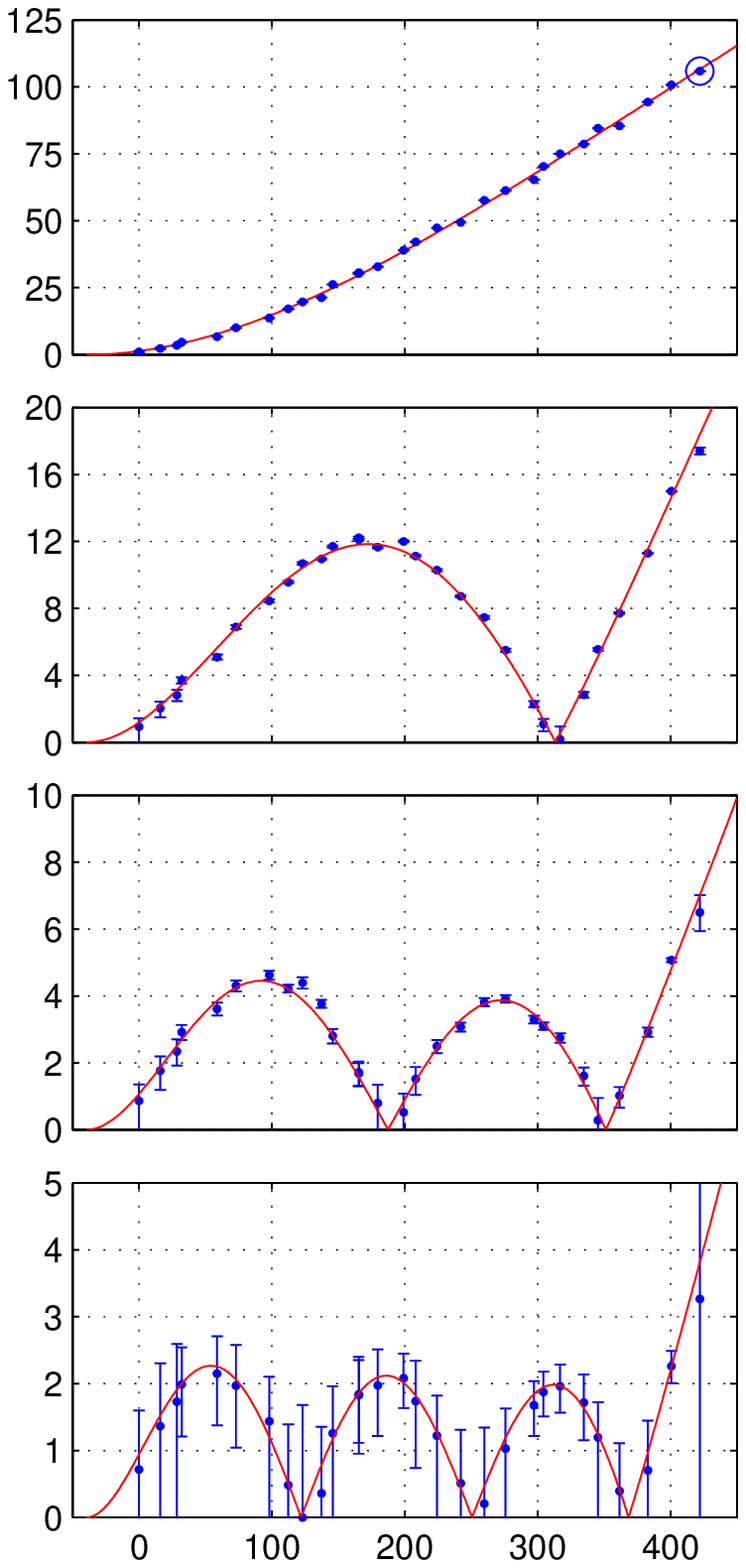}%
\drawat{-48mm}{1.30mm}{position $x/\SI{}{\mu m}$}%
\drawat{-81mm}{16mm}{\rotatebox{90}{$\sqrt{\left\langle\delta_{4}^{2}(x)\right\rangle}/\SI{}{pm}$}}%
\drawat{-81mm}{56mm}{\rotatebox{90}{$\sqrt{\left\langle\delta_{3}^{2}(x)\right\rangle}/\SI{}{pm}$}}%
\drawat{-81mm}{95mm}{\rotatebox{90}{$\sqrt{\left\langle\delta_{2}^{2}(x)\right\rangle}/\SI{}{pm}$}}%
\drawat{-81mm}{135mm}{\rotatebox{90}{$\sqrt{\left\langle\delta_{1}^{2}(x)\right\rangle}/\SI{}{pm}$}}%
\end{center}
\caption{Amplitude of thermal noise for the first 4 flexural modes along the cantilever. Errors bars correspond to the equivalent noise of the detection system in the bandwidth chosen around each resonance (a very conservative estimation of uncertainty). The simultaneous fit (red curves) of the 4 resonances with the normal modes shapes is excellent and leads to a precise measurement of the stiffness of the cantilever. The common calibration method for the spring constant considers only the value of $\delta_{1}^{2}(L)$, the circled point in the upper graph.}
\label{Fig:4fits}
\end{figure}

For a quantitative characterization of the shape of the modes we determine the rms amplitude of each resonance $\left\langle\delta_{n}^{2}(x)\right\rangle$ as a function of the position $x$, by integrating the PSD in a convenient frequency interval $2 \Delta f$ around each peak:
\begin{equation}
\left\langle \delta_{n}^{2}(x)\right\rangle=\int_{f_{n}-\Delta f}^{f_{n}+{\Delta f}} S_{d}(x,f){d\! f}
\end{equation}
This quantity is computed directly from the experimental spectrums, without any fitting process of the resonances. We anyway take care to subtract contribution of the background noise of the interferometer and to compensate for the finite integration range in frequency. Experimental data computed this way is plotted on Fig.~\ref{Fig:4fits}. Error bars correspond to the equivalent noise of the detection system in the bandwidth chosen around each resonance (a very conservative estimation of uncertainty).

According to equation \ref{EBsolution}, we should have
\begin{equation}
\left\langle \delta_{n}^{2}(x)\right\rangle=\left\langle d_{n}^{2}\right\rangle\left|\phi_{n}\left(\frac{x}{L}\right)\right|^{2}
\end{equation}
Using the expression of $\left\langle d_{n}^{2}\right\rangle$ (eq.~\ref{eq:equirepartition}), it is therefore possible to fit the data with 3 adjustable parameters: the length of the lever $L'$, the clamping position  $x_{0}$ and the static spring constant of the cantilever $k_{c}'$. We realize the fit on the rms amplitude \emph{simultaneously} on the 4 modes with the following functions:
\begin{equation}
\sqrt{\left\langle \delta_{n}^2(x)\right\rangle}=\sqrt{\frac{3}{\alpha_{n}^{4}}\frac{k_{B}T}{k_{c}'}}\left|\phi_{n}\left(\frac{x-x_0}{L'}\right)\right|     
\end{equation} 
where $n=1,...,4$. The red curves in Fig.~\ref{Fig:4fits} represent the fits of the four considered modes, in good agreement for all modes. The best fit values are: $L'=(450\pm5)\SI{}{\mu m}$ and $k_{c}'=(0.376\pm0.015)\SI{}{N/m}$. These length and stiffness are compatible with the values provided by the manufacturer ($L=(450\pm10)\SI{}{\mu m}$ and $k_{c}$ from $\SI{0.07}{N/m}$ to $\SI{0.4}{N/m}$).

It is worth noting that the accuracy of our instrument provides a precise measurement of the thermal noise driven deflection along the cantilever length and allows to verify the Euler-Bernoulli model for the micro-lever. Furthermore this multi-mode approach provides a more reliable way to estimate the spring constant of a cantilever with respect to the standard thermal noise calibration method, which is limited to the integral of the first mode only and just at the cantilever free end (circled point in Fig.~\ref{Fig:4fits}). Actually a precise measurement of stiffness could be obtained from the first 3 modes only, the presence of the nodes providing a favorable constraint to the fitting process. Moreover, the use of the interferometric set-up allows one to avoid the calibration of the segmented photodiode response as in the classical optical lever readout scheme: this step implies a contact between the AFM tip and a hard sample which is translated of a known amount, a process that may be undesirable to preserve the probe's sharpness or its coating. Our calibration method leaves only a small $\SI{4}{\%}$ uncertainty on the spring constant value for the cantilever used here (confidence interval corresponding to one standard deviation estimated during the linear least square fitting process).

\begin{figure}[ht]
\begin{center}
\includegraphics{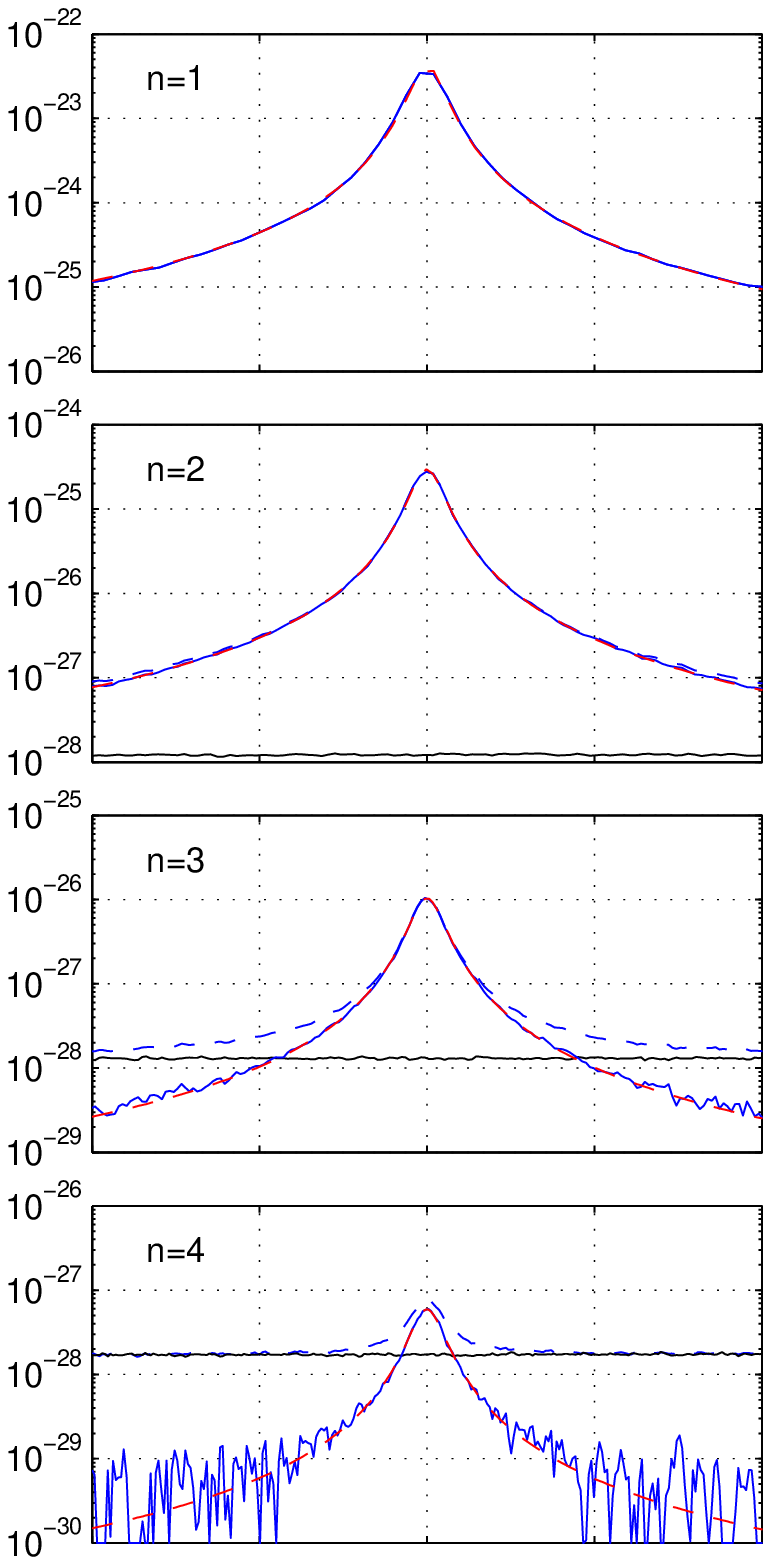}%
\drawat{-44mm}{-2mm}{frequency $f$}%
\drawat{-22mm}{3.5mm}{$+5 f_{n}/Q_{n}$}%
\drawat{-63mm}{3.5mm}{$-5 f_{n}/Q_{n}$}%
\drawat{-37mm}{3.5mm}{$f_{n}$}%
\drawat{-83mm}{11mm}{\rotatebox{90}{PSD $S_d/(\SI{}{m^2/Hz})$}}%
\drawat{-83mm}{51mm}{\rotatebox{90}{PSD $S_d/(\SI{}{m^2/Hz})$}}%
\drawat{-83mm}{90mm}{\rotatebox{90}{PSD $S_d/(\SI{}{m^2/Hz})$}}%
\drawat{-83mm}{130mm}{\rotatebox{90}{PSD $S_d/(\SI{}{m^2/Hz})$}}%
\end{center}
\caption{Power Spectrum Density (PSD) $S_{d}(x\approx L,f)$ of thermal noise induced deflection measured close to the free end of the cantilever as a function of frequency $f$ around each resonance $f_{n}$.  We subtract from the raw measurement (dashed blue) the background noise (black) to estimate the mechanical noise of the cantilever (blue), then we perform a fit with a simple damped harmonic oscillator model (dashed red). For each plot, the horizontal scale spans $\pm 10 f_{n}/Q_{n}$ around $f_{n}$.}
\label{Fig:Qfits}
\end{figure}

Experimental PSD curves show that resonances have a mode number dependent frequency width. This effect is mainly due to the viscous drag of the fluid, an point that is not considered in the Euler-Bernoulli framework. A simple model that accounts for this aspect, for each eigenmode, is a damped harmonic oscillator characterized by three parameters: the resonance frequency $f_{n}$, the elastic constant $k_{n}$ and the quality factor $Q_{n}$.
The PSD of the damped oscillator can be expressed as:
\begin{equation}
         S_{d}(f)= \frac{2k_{B}T}{k_{n}\pi f_{n}}\frac{Q_{n}}{{(1-u^2)}^2 Q_{n}^2+u^2}
\end{equation}
with $u=f/f_{n}$ the normalised frequency.
For each mode the PSD depends on the spatial coordinate along the cantilever length still according to $\left|\phi_{n}(x)\right|^2$. 
The $Q_{n}$ are evaluated from experimental data by a fit in the region around the peaks of each node (see Fig.~\ref{Fig:Qfits}). Those values are constant along the cantilever. Since the elastic constant $k_{n}$ and resonant frequency $f_{n}$ of each mode is known, the quality factor $Q_{n}$ is the only free parameter in these fits. In table~\ref{table} the $Q_{n}$ values of the first four modes are compared with the hydrodynamic predictions of the Sader model \cite{sader1998} that account for viscous effect in the fluid. These predictions were computed with tabulated values of silicon and gold for Young's modulus and density, and the physical dimensions of the cantilever (length, width and thickness) were tuned within the manufacturer tolerance to match the experimental observations. Note that the elastic modulus of gold and silicon are of the same order of magnitude, whereas the density of gold is about eight times that of silicon, therefore even a thin coating layer of $\SI{70}{nm}$ produces a mass increment of about $60{\%}$ that cannot be neglected in the evaluation of total mass.

A good agreement between the model and experiment is observed for the first three resonances, but at the highest frequency a deviation is observed: a lower dissipation than that foreseen by the Sader model. This behavior was already observed by Maali et al \cite{Maali2005}. 
This deviation is expected since the original Sader model neglects the 3D nature of the fluid flowing around the cantilever, an effect increasing with mode number \cite{sader1998}. In an extended model by Sader and co-workers \cite{Sader2007}, such a correction is also observed in the same direction as in our observation.

\section{Conclusions}

We presented in this paper a high precision measurement of thermal noise induced deflection of a soft rectangular micro-cantilever as a function of frequency and position along the mechanical beam. The 4 first flexural spatial modes could be studied without any external forcing. Their shapes are very well fitted by the Euler-Bernoulli model, and their vibration amplitudes accurately described within the Butt and Jaschke \cite{butt} framework. Furthermore this multi-mode approach provides an extension to the standard thermal noise calibration method for the spring constant of the cantilever, with a more robust estimation: it is obtained by a simultaneous fit on several modes, when the classic measurement is  limited to the integral of the first mode only and just at the cantilever free end. Here, the stiffness of our cantilever could be determined with only $\SI{4}{\%}$ uncertainty. Quality factors of resonances are also robustly extracted from the measurements and compare well to the Sader estimation \cite{sader1998}. 

The very simple geometry considered here, a rectangular AFM probe, is commonly used in many other applications. In fact single-clamped structures, similar to cantilevers, are often the basic elements of complex MEMS devices. Many physical, chemical and biological sensors, a large family of micro-devices, are based on cantilever shaped structures. They respond to an external change with a small, barely detectable, mechanical movement. The proposed calibration method would be directly applicable to these objects. The characterization of more complex structures (bi-dimensional for instance) such as arrays, double clamped elements, membranes and other micro-mechanical structures could also be performed, although it would also require a proper theoretical treatment to extract the relevant physical parameters from the measurement. The absence of external forcing (that may not be controlled or hard to characterize) thanks to the use of thermal noise and the great resolution of the interferometric setup make our approach a very promising tool for the mechanical characterization of MEMS. Even beyond such calibration, our setup proves to be a valuable tool to perform measurements of extremely small mechanical displacements with a high bandwidth.

\bigskip

{\bf Acknowledgements}
The authors thank F. Vittoz and F. Ropars for technical support, and S. Ciliberto, A. Petrosyan and J.P. Aim\'e for stimulating discussions. One of us (B.T.) thanks \'Ecole Normale Sup\'erieure de Lyon for invited professor position. This work has been partially supported by contract ANR-05-BLAN-0105-01 of the Agence Nationale de la Recherche in France.

\end{document}